\documentclass[12pt]{article}
\begin{document}
\setcounter{page}{1}
\def\theequation{\arabic{section}.\arabic{equation}}
\def\theequation{\thesection.\arabic{equation}}
\setcounter{section}{0}

\title{Solar proton burning, neutrino disintegration of the deuteron
and pep process in the relativistic field theory model of the
deuteron}

\author{A. N. Ivanov~\thanks{E--mail: ivanov@kph.tuwien.ac.at, Tel.:
+43--1--58801--14261, Fax: +43--1--58801--14299}~${
^\ddagger}$, H. Oberhummer~\thanks{E--mail: ohu@kph.tuwien.ac.at,
Tel.: +43--1--58801--14251, Fax: +43--1--58801--14299} ,
N. I. Troitskaya~\thanks{Permanent Address: State Technical
University, Department of Nuclear Physics, 195251 St. Petersburg,
Russian Federation} , M. Faber~\thanks{E--mail:
faber@kph.tuwien.ac.at, Tel.: +43--1--58801--14261, Fax:
+43--1--58801--14299}}

\date{\today}

\maketitle

\begin{center}
{\it Institut f\"ur Kernphysik, Technische Universit\"at Wien,\\
Wiedner Hauptstr. 8-10, A-1040 Vienna, Austria}
\end{center}

\begin{center}
\begin{abstract}
The astrophysical factor $S_{\rm pp}(0)$ for the solar proton burning
p + p $\to$ D + e$^+$ + $\nu_{\rm e}$ is recalculated in the
relativistic field theory model of the deuteron (RFMD). We obtain
$S_{\rm pp}(0) = 4.08 \times 10^{-25}\,{\rm MeV\,\rm b}$ which agrees
good with the recommended value $S_{\rm pp}(0) = 4.00 \times 10^{-25}\,{\rm
MeV\,\rm b}$. The amplitude of low--energy elastic proton--proton (pp)
scattering in the ${^1}{\rm S}_0$--state with the Coulomb repulsion
contributing to the amplitude of the solar proton burning is described
in terms of the S--wave scattering length and the effective range.
This takes away the problem pointed out by Bahcall and Kamionkowski
(Nucl. Phys. A625 (1997) 893) that in the RFMD one cannot describe
low--energy elastic pp scattering with the Coulomb repulsion in
agreement with low--energy nuclear phenomenology. The cross section
for the neutrino disintegration of the deuteron $\nu_{\rm e}$ + D
$\to$ e$^-$ + p + p is calculated with respect to $S_{\rm pp}(0)$ for
neutrino energies up to $E_{\nu_{\rm e}} \le 10\,{\rm MeV}$. The
results can be used for the analysis of the data which will be obtained in the
experiments planned by SNO.  The astrophysical factor $S_{\rm pep}(0)$
for the process p + e$^-$ + p $\to$ $\nu_{\rm e}$ + D (or
pep--process) is calculated relative to $S_{\rm pp}(0)$ in complete
agreement with the result obtained by Bahcall and May (ApJ. 155 (1969)
501).
\end{abstract}
\end{center}

\begin{center}
\end{center}

\newpage

\section{Introduction}
\setcounter{equation}{0} The weak nuclear process p + p $\to$ D +
e$^+$ + $\nu_{\rm e}$, the solar proton burning or proton--proton (pp)
fusion, plays an important role in Astrophysics [1,2]. It gives start
for the p--p chain of nucleosynthesis in the Sun and the
main--sequence stars [1,2]. In the Standard Solar Model (SSM) [3] the
total (or bolometric) luminosity of the Sun $L_{\odot} = (3.846\pm
0.008)\times 10^{26}\,{\rm W}$ is normalized to the astrophysical
factor $S_{\rm pp}(0)$ for pp fusion. The recommended value $S_{\rm
pp}(0) = 4.00\times 10^{-25}\,{\rm MeV b}$ [4] has been found by
averaging over the results obtained in the Potential model approach
(PMA) [5,6] and the Effective Field Theory (EFT) approach
[7,8]. However, as has been shown recently in Ref.[9] {\it the inverse
and forward helioseismic approach indicate the higher values of
$S_{\rm pp}(0)$ seem more favoured}, for example, $S_{\rm pp}(0) =
4.20\times 10^{-25}\,{\rm MeV b}$ and higher [9]. Of course,
accounting for the experimental errors the recommended value does not
contradict the result obtained in Ref.[9].

In Refs.[10-13] we have developed a relativistic field theory model of
the deuteron (RFMD). In turn, in Ref.[14] we have suggested a modified
version of the RFMD which is not well defined due to a violation of
Lorentz invariance of the effective four--nucleon interaction
describing N + N $\to$ N + N transitions. This violation has turned
out to be incompatible with a dominance of one--nucleon loop anomalies
which are Lorentz covariant. Thereby, the astrophysical factor $S_{\rm
pp}(0)$ calculated in the modified version of the RFMD [14] and enhanced by
a factor of 1.4 with respect to the recommended value [4] is not good
established.  This result demands the confirmation within the original
RFMD [10--13] by using the technique expounded in Ref.[13].

As has been shown in Ref.\,[12] the RFMD is motivated by QCD. The
deuteron appears in the nuclear phase of QCD as a neutron--proton
collective excitation -- a Cooper np--pair induced by a
phenomenological local four--nucleon interaction. Strong low--energy
interactions of the deuteron coupled to itself and other particles are
described in terms of one--nucleon loop exchanges. The one--nucleon
loop exchanges allow to transfer nuclear flavours from an initial to a
final nuclear state by a minimal way and to take into account
contributions of nucleon--loop anomalies determined completely by
one--nucleon loop diagrams. The dominance of contributions of
nucleon--loop anomalies has been justified in the large $N_C$
expansion, where $N_C$ is the number of quark colours [13]. Unlike the
PMA and the EFT approach the RFMD takes into account non--perturbative
contributions of high--energy (short--distance) fluctuations of
virtual nucleon ($N$) and anti--nucleon ($\bar{N}$) fields, $N\bar{N}$
fluctuations, in the form of one--nucleon loop anomalies. In accord
the analysis carried out in Refs.[15] nucleon--loop anomalies can be
interpreted as non--perturbative contributions of the nucleon Dirac
sea. The description of one--nucleon loop anomalies goes beyond the
scope of both the PMA and the EFT approach due to the absence in these
approaches anti--nucleon degrees of freedom related to the nucleon
Dirac sea. However, one should notice that in low--energy nuclear
physics the nucleon Dirac sea cannot be ignored fully [16]. For
example, high--energy $N\bar{N}$ fluctuations of the nucleon Dirac sea
polarized by the nuclear medium decrease the scalar nuclear density in
the nuclear interior of finite nuclei by 15$\%$ [16]. This effect has
been obtained within quantum field theoretic approaches in terms of
one--nucleon loop exchanges.

In this paper we revise the value of $S_{\rm pp}(0)$ obtained in
Ref.\,[14]. For this aim we apply the technique developed in the RFMD
[13] for the description of contributions of low--energy elastic
nucleon--nucleon scattering in the ${^1}{\rm S}_0$--state to
amplitudes of electromagnetic and weak nuclear processes. This
technique implies the summation of an infinite series of one--nucleon
loop diagrams and the evaluation of the result of the summation in
leading order in large $N_C$ expansion [13]. The application of this
method to the evaluation of the cross sections for the anti--neutrino
disintegration of the deuteron induced by charged $\bar{\nu}_{\rm e}$
+ D $\to$ e$^+$ + n + n and neutral $\bar{\nu}_{\rm e}$ + D $\to$
$\bar{\nu}_{\rm e}$ + n + p weak currents gave the results  agreeing
good with the experimental data. The reaction $\bar{\nu}_{\rm e}$ + D
$\to$ e$^+$ + n + n is, in the sense of charge independence of weak
interaction strength, equivalent to the reaction p + p $\to$ D + e$^+$
+ $\nu_{\rm e}$. Therefore, the application of the same technique to
the description of the reaction p + p $\to$ D + e$^+$ + $\nu_{\rm e}$
should give a result of a good confidence level.

The paper is organized as follows. In Sect.\,2 we evaluate the
amplitude of the solar proton burning. We show that the contribution
of low--energy elastic pp scattering in the ${^1}{\rm S}_0$--state
with the Coulomb repulsion is described in agreement with low--energy
nuclear phenomenology in terms of the S--wave scattering length and
the effective range. This takes away the problem pointed out by
Bahcall and Kamionkowski [17] that in the RFMD one cannot describe
low--energy elastic pp scattering with the Coulomb repulsion in
agreement with low--energy nuclear phenomenology. In Sect.\,3 we
evaluate the astrophysical factor for the solar proton burning and
obtain the value $S_{\rm pp}(0) = 4.08\times 10^{-25}\,{\rm MeV\, b}$
agreeing good with the recommended one $S_{\rm pp}(0) = 4.00\times
10^{-25}\,{\rm MeV\, b}$.  In Sect.\,4 we evaluate the cross section for
the neutrino disintegration of the deuteron $\nu_{\rm e}$ + D $\to$
e$^-$ + p + p caused by the charged weak current with respect to
$S_{\rm pp}(0)$. In Sect.\,5 we adduce the evaluation of the
astrophysical factor $S_{\rm pep}(0)$ of the reaction p + e$^-$ + p
$\to$ D + $\nu_{\rm e}$ or pep--process relative to $S_{\rm
pp}(0)$. In the Conclusion we discuss the obtained results.

\section{Amplitude of solar proton burning and low--energy elastic 
proton--proton scattering} 
\setcounter{equation}{0}

For the description of low--energy transitions N + N $\to$ N + N in
the reactions n + p $\to$ D + $\gamma$, $\gamma$ + D $\to$ n + p,
$\bar{\nu}_{\rm e}$ + D $\to$ e$^+$ + n + n and p + p $\to$ D + e$^+$
+ $\nu_{\rm e}$, where nucleons are in the ${^1}{\rm S}_0$--state, we
apply the effective local four--nucleon interactions [11--13]:
\begin{eqnarray}\label{label2.1}
&&{\cal L}^{\rm NN \to NN}_{\rm eff}(x)=G_{\rm \pi
NN}\,\{[\bar{n}(x)\gamma_{\mu}
\gamma^5 p^c(x)][\bar{p^c}(x)\gamma^{\mu}\gamma^5 n(x)]\nonumber\\
&&+\frac{1}{2}\,
[\bar{n}(x)\gamma_{\mu} \gamma^5 n^c(x)][\bar{n^c}(x)\gamma^{\mu}\gamma^5
n(x)] +
\frac{1}{2}\,[\bar{p}(x)\gamma_{\mu} \gamma^5 p^c(x)]
[\bar{p^c}(x)\gamma^{\mu}\gamma^5 p(x)]\nonumber\\
&&+ (\gamma_{\mu}\gamma^5 \otimes \gamma^{\mu}\gamma^5 \to \gamma^5 \otimes
\gamma^5)\},
\end{eqnarray}
where $n(x)$ and $p(x)$ are the operators of the neutron and the
proton interpolating fields, $n^c(x) = C \bar{n}^T(x)$ and so on, then
$C$ is a charge conjugation matrix and $T$ is a transposition.  The
effective coupling constant $G_{\rm \pi NN}$ is defined by [11--13]
\begin{eqnarray}\label{label2.2}
G_{\rm \pi NN} = \frac{g^2_{\rm \pi NN}}{4M^2_{\pi}} - \frac{2\pi a_{\rm
np}}{M_{\rm N}} = 3.27\times 10^{-3}\,{\rm MeV}^{-2},
\end{eqnarray}
where $g_{\rm \pi NN}= 13.4$ is the coupling constant of the
${\rm \pi NN}$ interaction, $M_{\pi}=135\,{\rm MeV}$ is the pion
mass, $M_{\rm p} = M_{\rm n} = M_{\rm N} = 940\,{\rm MeV}$ is the mass
of the proton and the neutron neglecting the electromagnetic mass
difference, which is taken into account only for the calculation of
the phase volumes of the final states of the reactions p + p $\to$ D +
e$^+$ + $\nu_{\rm e}$, $\nu_{\rm e}$ + D $\to$ e$^-$ + p + p and p +
e$^-$ + p $\to$ D + $\nu_{\rm e}$, and $a_{\rm np} = (-23.75\pm
0.01)\,{\rm fm}$ is the S--wave scattering length of np scattering in
the ${^1}{\rm S}_0$--state.

The effective Lagrangian for the low--energy nuclear transition p + p
$\to$ D + e$^+$ + $\nu_{\rm e}$ has been calculated in Ref.\,[12] and reads
\begin{eqnarray}\label{label2.3}
{\cal L}_{\rm pp\to D e^+ \nu_{\rm e}}(x) = - i g_{\rm A}G_{\rm \pi
NN}M_{\rm N}\frac{G_{\rm V}}{\sqrt{2}}\frac{3g_{\rm
V}}{4\pi^2}\,D^{\dagger}_{\mu}(x)\,[\bar{p^c}(x)\gamma^5
p(x)]\,[\bar{\psi}_{\nu_{\rm e}}(x)\gamma^{\mu}(1 - \gamma^5) \psi_{\rm
e}(x)].
\end{eqnarray}
where $G_{\rm V} = G_{\rm F}\,\cos \vartheta_C$ with $G_{\rm F} =
1.166\,\times\,10^{-11}\,{\rm MeV}^{-2}$ and $\vartheta_C$ are the
Fermi weak coupling constant and the Cabibbo angle $\cos \vartheta_C =
0.975$, $g_{\rm A} = 1.2670 \pm 0.0035$ [18] and $g_{\rm V}$ is a
phenomenological coupling constant of the RFMD related to the electric
quadrupole moment of the deuteron $Q_{\rm D} = 0.286\,{\rm fm}^2$
[11]: $g^2_{\rm V} = 2\pi^2 Q_{\rm D}M^2_{\rm N}$. Then, $D_{\mu}(x)$,
$\psi_{\nu_{\rm e}}(x)$ $\psi_{\rm e}(x)$ are the interpolating
fields of the deuteron and leptonic pair, respectively.

The effective Lagrangian Eq.(\ref{label2.3}) defines the effective
vertex of the low--energy nuclear transition  p + p
$\to$ D + e$^+$ + $\nu_{\rm e}$ 
\begin{eqnarray}\label{label2.4}
i{\cal M}({\rm p} + {\rm p} \to {\rm D} + {\rm e}^+ + \nu_{e})&=&
\,G_{\rm V}\,g_{\rm A} M_{\rm N}\,G_{\rm \pi NN}\,\frac{3g_{\rm V}}{4\pi^2}\,
e^*_{\mu}(k_{\rm D})\,[\bar{u}(k_{\nu_{\rm
e}})\gamma^{\mu} (1-\gamma^5) v(k_{\rm e^+})]\nonumber\\
&&\times\,[\bar{u^c}(p_2) \gamma^5
u(p_1)],
\end{eqnarray}
where $e^*_{\mu}(k_{\rm D})$ is a 4--vector of a polarization of the
deuteron, $u(k_{\nu_{\rm e}})$, $v(k_{\rm e^+})$,
$u(p_2)$ and $u(p_1)$ are the Dirac bispinors of neutrino,
positron  and two protons, respectively.

In order to evaluate the contribution of low--energy elastic pp
scattering we have to determine the effective vertex of the p + p
$\to$ p + p transition accounting for the Coulomb repulsion between
the protons. For this aim we suggest to use the effective local
four--nucleon interaction Eq.(\ref{label2.1}) and take into account
the Coulomb repulsion in terms of the explicit Coulomb wave function
of the protons. This yields
\begin{eqnarray}\label{label2.5}
V_{\rm pp \to pp}(k',k) = G_{\rm \pi NN}\,\psi^*_{\rm pp}(k'\,)\,
[\bar{u}(p'_2) \gamma^5 u^c(p'_1)]\,[\bar{u^c}(p_2) \gamma^5
u(p_1)]\,\psi_{\rm pp}(k),
\end{eqnarray}
where $\psi_{\rm pp}(k)$ and $\psi^*_{\rm pp}(k'\,)$ are the explicit
Coulomb wave functions of the relative movement of the protons taken
at zero relative radius, and $k$ and $k'$ are relative 3--momenta
of the protons $\vec{k} = (\vec{p}_1 - \vec{p}_2)/2$ and
$\vec{k}^{\,\prime} = (\vec{p}^{\;\prime}_1  - \vec{p}^{\;\prime}_2 )/2$ in the
initial and final states. The explicit form of $\psi_{\rm pp}(k)$ we
take following Kong and Ravndal [8] (see also [19])
\begin{eqnarray}\label{label2.6}
\psi_{\rm pp}(k) = e^{\textstyle - \pi/4k r_C}\,\Gamma\Bigg(1 +
\frac{i}{2k r_C}\Bigg),
\end{eqnarray}
where $r_C = 1/M_{\rm N}\alpha = 28.82\,{\rm fm}$ and $\alpha = 1/137$
are the Bohr radius of a proton and the fine structure constant. The
squared value of the modulo of $\psi_{\rm pp}(k)$ is given by
\begin{eqnarray}\label{label2.7}
|\psi_{\rm pp}(k)|^2 = C^2_0(k) = \frac{\pi}{k r_C}\,
\frac{1}{\displaystyle e^{\textstyle \pi/k r_C} - 1},
\end{eqnarray}
where $C_0(k)$ is the Gamow penetration factor [1,2,19].  We would
like to emphasize that the wave function Eq.(\ref{label2.6}) is
defined only by a regular solution of the Schr\"odinger equation for
the pure Coulomb potential [19].

By taking into account the contribution of the Coulomb wave function
and summing up an infinite series of one--proton loop diagrams the
amplitude of the solar proton burning can be written in the form
\begin{eqnarray}\label{label2.8}
\hspace{-0.5in}&&i{\cal M}({\rm p} + {\rm p} \to {\rm D} + {\rm e}^+ +
\nu_{e}) = G_{\rm V}\,g_{\rm A} M_{\rm N}\,G_{\rm \pi
NN}\,\frac{3g_{\rm V}}{4\pi^2}\, e^*_{\mu}(k_{\rm
D})\,[\bar{u}(k_{\nu_{\rm e}})\gamma^{\mu} (1-\gamma^5) v(k_{\rm
e^+})]\,{\cal F}^{\rm e}_{\rm pp}\nonumber\\
\hspace{-0.5in}&&\times\,\frac{[\bar{u^c}(p_2) \gamma^5 u(p_1)]\,\psi_{\rm
pp}(k)}{\displaystyle 1 + \frac{G_{\rm \pi NN}}{16\pi^2}\int
\frac{d^4p}{\pi^2i}\,|\psi_{\rm pp}(|\vec{p} + \vec{Q}\,|)|^2 {\rm
tr}\Bigg\{\gamma^5 \frac{1}{M_{\rm N} - \hat{p} - \hat{P} -
\hat{Q}}\gamma^5 \frac{1}{M_{\rm N} - \hat{p} - \hat{Q}}\Bigg\}}.
\end{eqnarray}
where $P = p_1 + p_2 = (2\sqrt{k^2 + M^2_{\rm N}}, \vec{0}\,)$ is the
4--momentum of the pp--pair in the center of mass frame; $Q =a\,P +
b\,K = a\,(p_1 + p_2) + b\,(p_1 - p_2)$ is an arbitrary shift of
virtual momentum with arbitrary parameters $a$ and $b$, and in the
center of mass frame $K = p_1 - p_2 = (0,2\,\vec{k}\,)$ [14]. The
parameters $a$ and $b$ can be functions of $k$. The factor ${\cal
F}^{\rm e}_{\rm pp}$ describes the overlap of the Coulomb and strong
interactions [10]. It is analogous the overlap integral in the PMA
[5]. We calculate this factor below.

The evaluation of the momentum integral runs the way expounded in
[14]. Keep only leading contributions in the large $N_C$ expansion
[13,14] we obtain
\begin{eqnarray}\label{label2.9}
&&\int \frac{d^4p}{\pi^2i}\,|\psi_{\rm
pp}(|\vec{p} + \vec{Q}\,|)|^2 {\rm tr}\Bigg\{\gamma^5 \frac{1}{M_{\rm N}
- \hat{p} - \hat{P} - \hat{Q}}\gamma^5 \frac{1}{M_{\rm N} - \hat{p} -
\hat{Q}}\Bigg\} =\nonumber\\
&&= - 8\, a\,(a + 1)\,M^2_{\rm N} + 8\,(b^2 - a\,(a +
1))\,k^2 - i\,8\pi\,M_{\rm N}\,k\,|\psi_{\rm pp}(k)|^2 = \nonumber\\
&&= - 8\, a\,(a + 1)\,M^2_{\rm N} + 8\,(b^2 - a\,(a +
1))\,k^2 - i\,8\pi\,M_{\rm N}\,k\,C^2_0(k).
\end{eqnarray}
Substituting Eq.(\ref{label2.9}) in Eq.(\ref{label2.8}) we get 
\begin{eqnarray}\label{label2.10}
\hspace{-0.7in}&& i{\cal M}({\rm p} + {\rm p} \to {\rm D} + {\rm e}^+
+ \nu_{e}) = G_{\rm V}\,g_{\rm A} M_{\rm N}\,G_{\rm \pi
NN}\,\frac{3g_{\rm V}}{4\pi^2}\,{\cal F}^{\rm e}_{\rm pp}\nonumber\\
\hspace{-0.7in}&&\times \, e^*_{\mu}(k_{\rm D})\,[\bar{u}(k_{\nu_{\rm
e}})\gamma^{\mu} (1-\gamma^5) v(k_{\rm e^+})]\,[\bar{u^c}(p_2)
\gamma^5 u(p_1)]\,e^{\textstyle - \pi/4k
r_C}\,\Gamma\Bigg(1 + \frac{i}{2k r_C}\Bigg)\nonumber\\
\hspace{-0.7in}&&\Bigg[ 1 - a(a+1) \frac{G_{\rm
\pi NN}}{2\pi^2}\,M^2_{\rm N} + \frac{G_{\rm \pi NN}}{2\pi^2}\,(b^2 -
a\,(a + 1))\,k^2 - i\,\frac{G_{\rm \pi NN}M_{\rm
N}}{2\pi}\,k\,C^2_0(k)\Bigg]^{-1}\!\!\!.
\end{eqnarray}
In order to reconcile the contribution of low--energy elastic pp
scattering with low--energy nuclear phenomenology [19] we should make
a few changes. For this aim we should rewrite Eq.(\ref{label2.10}) in more
convenient form
\begin{eqnarray}\label{label2.11}
\hspace{-0.7in}&& i{\cal M}({\rm p} + {\rm p} \to {\rm D} + {\rm e}^+
+ \nu_{e}) = G_{\rm V}\,g_{\rm A} M_{\rm N}\,G_{\rm \pi
NN}\,\frac{3g_{\rm V}}{4\pi^2}\,{\cal F}^{\rm e}_{\rm pp}\nonumber\\
\hspace{-0.7in}&&\times \, e^*_{\mu}(k_{\rm D})\,[\bar{u}(k_{\nu_{\rm
e}})\gamma^{\mu} (1-\gamma^5) v(k_{\rm e^+})]\,[\bar{u^c}(p_2)
\gamma^5 u(p_1)]\,e^{\textstyle i\sigma_0(k)}\,C_0(k)\nonumber\\
\hspace{-0.7in}&&\Bigg[ 1 - a(a+1) \frac{G_{\rm
\pi NN}}{2\pi^2}\,M^2_{\rm N} + \frac{G_{\rm \pi NN}}{2\pi^2}\,(b^2 -
a\,(a + 1))\,k^2 - i\,\frac{G_{\rm \pi NN}M_{\rm
N}}{2\pi}\,k\,C^2_0(k)\Bigg]^{-1}\!\!\!.
\end{eqnarray}
We have denoted
\begin{eqnarray}\label{label2.12}
e^{\textstyle - \pi/4k r_C}\,\Gamma\Bigg(1 + \frac{i}{2k r_C}\Bigg)
= e^{\textstyle i\sigma_0(k)}\,C_0(k)\;,\;
\sigma_0(k)&=&{\rm arg}\,\Gamma\Bigg(1 + \frac{i}{2k r_C}\Bigg),
\end{eqnarray}
where $\sigma_0(k)$ is a pure Coulomb phase shift.

Now, let us rewrite the denominator of the amplitude
 Eq.(\ref{label2.11}) in the equivalent form
\begin{eqnarray}\label{label2.13}
&& \Bigg\{\cos\sigma_0(k)\Bigg[1 - a(a+1) \frac{G_{\rm \pi
NN}}{2\pi^2}\,M^2_{\rm N} + \frac{G_{\rm \pi NN}}{2\pi^2}\,(b^2 -
a\,(a + 1))\,k^2\Bigg]\nonumber\\
&& - \sin\sigma_0(k)\,\frac{G_{\rm \pi NN}M_{\rm
N}}{2\pi}\,k\,C^2_0(k)\Bigg\}- i\,\Bigg\{\cos\sigma_0(k)\,\frac{G_{\rm
\pi NN}M_{\rm N}}{2\pi}\,k\,C^2_0(k)\nonumber\\
&&  + \sin\sigma_0(k)\Bigg[1 - a(a+1) \frac{G_{\rm \pi
NN}}{2\pi^2}\,M^2_{\rm N} + \frac{G_{\rm \pi NN}}{2\pi^2}\,(b^2 -
a\,(a + 1))\,k^2\Bigg]\Bigg\}=\nonumber\\
&&= \frac{1}{Z}\Bigg[1 -
\frac{1}{2}\,a^{\rm e}_{\rm pp} r^{\rm e}_{\rm pp}k^2 + \frac{a^{\rm
e}_{\rm pp}}{r_C}\,h(2 k r_C) + i\,a^{\rm e}_{\rm pp}\,k\,C^2_0(k)\Bigg],
\end{eqnarray}
where we have denoted
\begin{eqnarray}\label{label2.14}
\hspace{-0.3in}&& \frac{1}{Z}\Bigg[1 - \frac{1}{2}\,a^{\rm e}_{\rm pp}
r^{\rm e}_{\rm pp}k^2 + \frac{a^{\rm e}_{\rm pp}}{r_C}\,h(2 k r_C)
\Bigg]= - \sin\sigma_0(k)\,\frac{G_{\rm \pi NN}M_{\rm
N}}{2\pi}\,k\,C^2_0(k)\nonumber\\
\hspace{-0.3in}&& +  \cos\sigma_0(k)\Bigg[1 - a(a+1) \frac{G_{\rm
\pi NN}}{2\pi^2}\,M^2_{\rm N} + \frac{G_{\rm \pi NN}}{2\pi^2}\,(b^2 -
a\,(a + 1))\,k^2\Bigg],\nonumber\\
\hspace{-0.3in}&& - \frac{1}{Z}\,a^{\rm e}_{\rm
pp}\,k\,C^2_0(k) =\cos\sigma_0(k)\frac{G_{\rm \pi NN}M_{\rm
N}}{2\pi}\,k\,C^2_0(k)\nonumber\\
\hspace{-0.3in}&& + \sin\sigma_0(k)\Bigg[1 - a(a+1) \frac{G_{\rm \pi
NN}}{2\pi^2}\,M^2_{\rm N} + \frac{G_{\rm \pi NN}}{2\pi^2}\,(b^2 -
a\,(a + 1))\,k^2\Bigg].
\end{eqnarray}
Here $Z$ is a constant which will be removed the renormalization of
the wave functions of the protons, $a^{\rm e}_{\rm pp} = ( - 7.8196\pm
0.0026)\,{\rm fm}$  and $r^{\rm e}_{\rm pp} = 2.790\pm 0.014\,{\rm
fm}$ [20] are the S--wave scattering length and the effective range of
pp scattering in the ${^1}{\rm S}_0$--state with the Coulomb
repulsion, and $h(2 k r_C)$ is defined by [19]
\begin{eqnarray}\label{label2.15}
h(2 k r_C) = - \gamma + {\ell n}(2 k r_C) +
\sum^{\infty}_{n=1}\frac{1}{n(1 + 4n^2k^2r^2_C)}.
\end{eqnarray}
The validity of the relations Eq.(\ref{label2.14}) assumes the
dependence of parameters $a$ and $b$ on the relative momentum $k$.

After the changes Eq.(\ref{label2.11})--Eq.(\ref{label2.14}) the
amplitude Eq.(\ref{label2.10}) takes the form
\begin{eqnarray}\label{label2.16}
\hspace{-0.2in}&& i{\cal M}({\rm p} + {\rm p} \to {\rm D} + 
{\rm e}^+ + \nu_{e}) =
G_{\rm V}\,g_{\rm A} M_{\rm N}\,G_{\rm \pi NN}\,\frac{3g_{\rm
V}}{4\pi^2}\,e^*_{\mu}(k_{\rm D})\,[\bar{u}(k_{\nu_{\rm
e}})\gamma^{\mu} (1-\gamma^5) v(k_{\rm e^+})]\,{\cal F}^{\rm e}_{\rm
pp}\nonumber\\ \hspace{-0.2in}&&\times \,\frac{C_0(k)}{\displaystyle1 -
\frac{1}{2}\,a^{\rm e}_{\rm pp} r^{\rm e}_{\rm pp}k^2 + \frac{a^{\rm
e}_{\rm pp}}{r_C}\,h(2 k r_C) + i\,a^{\rm e}_{\rm pp}\,k\,C^2_0(k)
}\,Z\,[\bar{u^c}(p_2) \gamma^5 u(p_1)].
\end{eqnarray}
Following [14] and renormalizing the wave functions of the protons
$\sqrt{Z}u(p_2) \to u(p_2)$ and
$\sqrt{Z}u(p_1) \to u(p_1)$ we obtain the amplitude of
the solar proton burning 
\begin{eqnarray}\label{label2.17}
\hspace{-0.2in}&& i{\cal M}({\rm p} + {\rm p} \to {\rm D} 
+ {\rm e}^+ + \nu_{e}) =
G_{\rm V}\,g_{\rm A} M_{\rm N}\,G_{\rm \pi NN}\,\frac{3g_{\rm
V}}{4\pi^2}\,e^*_{\mu}(k_{\rm D})\,[\bar{u}(k_{\nu_{\rm
e}})\gamma^{\mu} (1-\gamma^5) v(k_{\rm e^+})]\,{\cal F}^{\rm e}_{\rm
pp}\nonumber\\ \hspace{-0.2in}&&\times \,
\frac{\displaystyle C_0(k)}{\displaystyle1 -
\frac{1}{2}\,a^{\rm e}_{\rm pp} r^{\rm e}_{\rm pp}k^2 + \frac{a^{\rm
e}_{\rm pp}}{r_C}\,h(2 k r_C) + i\,a^{\rm e}_{\rm pp}\,k\,C^2_0(k)
}\,[\bar{u^c}(p_2) \gamma^5 u(p_1)]\,F_{\rm D}(k^2),
\end{eqnarray}
where we have introduced too an universal form factor [14]
\begin{eqnarray}\label{label2.18}
F_{\rm D}(k^2) =  \frac{1}{1 + r^2_{\rm D}k^2}
\end{eqnarray}
describing a spatial smearing of the deuteron coupled to the NN system
in the ${^1}{\rm S}_0$--state at low energies; $r_{\rm D} =
1/\sqrt{\varepsilon_{\rm D}M_{\rm N}} = 4.315\,{\rm fm}$ is the radius
of the deuteron and $\varepsilon_{\rm D} = 2.225\,{\rm MeV}$ is the
binding energy of the deuteron. 

The real part of the denominator of the amplitude Eq.(\ref{label2.17})
is in complete agreement with a phenomenological relation [19]
\begin{eqnarray}\label{label2.19}
{\rm ctg}\delta^{\rm e}_{\rm pp}(k) = \frac{1}{\displaystyle
C^2_0(k)\,k}\,\Bigg[ - \frac{1}{a^{\rm e}_{\rm pp}} +  \frac{1}{2}\,r^{\rm
e}_{\rm pp}k^2 -  \frac{1}{r_{\rm C}}\,h(2 k r_{\rm C})\Bigg],
\end{eqnarray}
describing the phase shift $\delta^{\rm e}_{\rm pp}(k)$ of low--energy
elastic pp scattering in terms of the S--wave scattering length
$a^{\rm e}_{\rm pp}$ and the effective range $r^{\rm e}_{\rm pp}$. As
has been pointed out [19] the expansion Eq.(\ref{label2.19}) is valid
up to $T_{\rm pp}\le 10\,{\rm MeV}$, where $T_{\rm pp} = k^2/M_{\rm
N}$ is a kinetic energy of the relative movement of the protons.

Thus, we argue that the contribution of low--energy elastic pp
scattering to the amplitude of the solar proton burning is described
in agreement with low--energy nuclear phenomenology in terms of the
S--wave scattering length $a^{\rm e}_{\rm pp}$ and the effective range
$r^{\rm e}_{\rm pp}$ taken from the experimental data [20]. This takes
away the problem pointed out by Bahcall and Kamionkowski [17] that in
the RFMD with the local four--nucleon interaction given by
Eq.(\ref{label2.1}) one cannot describe low--energy elastic pp
scattering with the Coulomb repulsion in agreement with low--energy
nuclear phenomenology.

Now let us proceed to the evaluation of ${\cal F}^{\rm e}_{\rm
pp}$. For this aim we should write down the matrix element of the
transition p + p $\to$ D + e$^+$ + $\nu_{\rm e}$ with the Coulomb
repulsion. The required matrix element has been derived in
Refs.[11,14] and reads
\begin{eqnarray}\label{label2.20}
\hspace{-0.3in}&&i{\cal M}_C({\rm p} + {\rm p} \to {\rm D} + {\rm e}^+ +
\nu_{e}) =\nonumber\\
\hspace{-0.3in}&&= G_{\rm V}\,g_{\rm A} M_{\rm
N}\,G_{\rm \pi NN}\,\frac{3g_{\rm
V}}{4\pi^2}\,C_0(k)\,[\bar{u}(k_{\nu_{\rm e}})\gamma^{\mu}
(1-\gamma^5) v(k_{\rm e^+})]\,e^*_{\mu}(k_{\rm
D})\,\nonumber\\
\hspace{-0.3in}&&\times\,\{- [\bar{u^c}(p_2)\gamma_{\alpha} \gamma^5
u(p_1)]{\cal J}^{\alpha\mu\nu}_C(k_{\rm D}, k_{\ell}) -
[\bar{u^c}(p_2)\gamma^5 u(p_1)]\,{\cal J}^{\mu\nu}_C(k_{\rm D},
k_{\ell})\},
\end{eqnarray}
where $k_{\rm D}$ and $k_{\ell}$ are 4--momenta of the deuteron and
the leptonic pair, respectively. The structure functions ${\cal
J}^{\alpha\mu\nu}(k_{\rm D}, k_{\ell})$ and ${\cal J}^{\mu\nu}(k_{\rm
D}, k_{\ell})$ are determined by [11,14]
\begin{eqnarray}\label{label2.21}
&&{\cal J}^{\alpha\mu\nu}_C(k_{\rm D}, k_{\ell}) =
\int\frac{d^4p}{\pi^2i}\,e^{\textstyle - \pi/4|\vec{q}\,|
r_C}\,\Gamma\Bigg(1 - \frac{i}{2 |\vec{q}\,| r_C}\Bigg)\nonumber\\
&&\times\,{\rm tr} \Bigg\{\gamma^{\alpha}\gamma^5\frac{1}{M_{\rm N} - \hat{p}
+ \hat{k}_{\rm D}}\gamma^{\mu}\frac{1}{M_{\rm N} -
\hat{p}}\gamma^{\nu}\gamma^5 \frac{1}{M_{\rm N} - \hat{p} -
\hat{k}_{\ell}}\Bigg\},\nonumber\\ 
&&{\cal J}^{\mu\nu}_C(k_{\rm D},k_{\ell}) =
\int\frac{d^4p}{\pi^2i}\,e^{\textstyle - \pi/4|\vec{q}\,|
r_C}\,\Gamma\Bigg(1 - \frac{i}{2 |\vec{q}\,| r_C}\Bigg)\nonumber\\
&&\times\,{\rm tr} \Bigg\{\gamma^5\frac{1}{M_{\rm N} - \hat{p}
+ \hat{k}_{\rm D}}\gamma^{\mu}\frac{1}{M_{\rm N} -
\hat{p}}\gamma^{\nu}\gamma^5 \frac{1}{M_{\rm N} - \hat{p} -
\hat{k}_{\ell}}\Bigg\},
\end{eqnarray}
where $\vec{q} = \vec{p} + (\vec{k}_{\ell} - \vec{k}_{\rm D})/2$.

For the subsequent analysis it is convenient to represent the
structure functions in the form of two terms
\begin{eqnarray}\label{label2.22}
{\cal J}^{\alpha\mu\nu}_C(k_{\rm D}, k_{\ell})&=&{\cal
J}^{\alpha\mu\nu}_{SS}(k_{\rm D}, k_{\ell}) + {\cal
J}^{\alpha\mu\nu}_{SC}(k_{\rm D}, k_{\ell}),\nonumber\\
{\cal J}^{\mu\nu}_C(k_{\rm D}, k_{\ell})&=&{\cal
J}^{\mu\nu}_{SS}(k_{\rm D}, k_{\ell}) + {\cal
J}^{\mu\nu}_{SC}(k_{\rm D}, k_{\ell}).
\end{eqnarray}
The decomposition is caused by the change
\begin{eqnarray}\label{label2.23}
e^{\textstyle - \pi/4|\vec{q}\,| r_C}\,\Gamma\Bigg(1 - \frac{i}{2
|\vec{q}\,| r_C}\Bigg) = 1 + \Bigg[e^{\textstyle - \pi/4|\vec{q}\,|
r_C}\,\Gamma\Bigg(1 - \frac{i}{2 |\vec{q}\,| r_C}\Bigg) - 1\Bigg],
\end{eqnarray}
where the first term gives the contribution to the $SS$ part of the
structure functions defined by strong interactions only, while the
second one vanishes at $r_C \to \infty$ ( or $\alpha \to 0$) and
describes the contribution to the $SC$ part of the structure functions
caused by both strong and Coulomb interactions.

The procedure of the evaluation of the structure functions
Eq.(\ref{label2.21}) and Eq.(\ref{label2.22}) has been described in
detail in Ref.[11,14]. Following this procedure we obtain ${\cal
F}^{\rm e}_{\rm pp}$ in the form
\begin{eqnarray}\label{label2.24}
\hspace{-0.3in}&&{\cal F}^{\rm e}_{\rm pp} =\nonumber\\
\hspace{-0.3in}&&= 1 +
\frac{32}{9}\int\limits^{\infty}_0 dp\,p^2\,\Bigg[e^{\textstyle -
\pi/4p r_C}\,\Gamma\Bigg(1 - \frac{i}{2pr_C}\Bigg)-
1\Bigg]\Bigg[\frac{M^2_{\rm N}}{(M^2_{\rm N} + p^2)^{5/2}} -
\frac{7}{16}\,\frac{1}{(M^2_{\rm N} + p^2)^{3/2}}\Bigg]\nonumber\\
\hspace{-0.3in}&&= 1 + \frac{32}{9}\int\limits^{\infty}_0
dv\,v^2\,\Bigg[e^{\textstyle - \alpha\pi/4v}\,\Gamma\Bigg(1 -
\frac{i\alpha}{2v}\Bigg)- 1\Bigg]\Bigg[\frac{1}{(1 + v^2)^{5/2}} -
\frac{7}{16}\,\frac{1}{(1 + v^2)^{3/2}}\Bigg].
\end{eqnarray}
The integral can be estimated perturbatively. The result reads
\begin{eqnarray}\label{label2.25}
{\cal F}^{\rm e}_{\rm pp} = 1 +
\alpha\,\Bigg(\frac{5\pi}{54} - i\,\frac{5\gamma}{27}\Bigg) + O(\alpha^2).
\end{eqnarray}
The numerical value of $|{\cal F}^{\rm e}_{\rm pp}|^2$ is
\begin{eqnarray}\label{label2.26}
|{\cal F}^{\rm e}_{\rm pp}|^2 = 1 + \alpha\,\frac{5\pi}{27} +
 O(\alpha^2) = 1 +(4.25 \times 10^{-3}) \simeq 1.
\end{eqnarray}
The contribution of the Coulomb field Eq.(\ref{label2.26}) inside the
one--nucleon loop diagrams is found small. This is because of the
integrals are concentrated around virtual momenta of order of $M_{\rm
N}$ which is of order $M_{\rm N} \sim N_C$ in the large $N_C$
expansion [12]. For the calculation of the astrophysical factor
$S_{\rm pp}(0)$ we can set ${\cal F}^{\rm e}_{\rm pp} = 1$.

\section{Astrophysical factor for  solar proton burning}
\setcounter{equation}{0}

The amplitude Eq.(\ref{label2.17}) squared, averaged over polarizations of
protons and summed over polarizations of final particles  reads 
\begin{eqnarray}\label{label3.1}
\hspace{-0.5in}&&\overline{|{\cal M}({\rm p} + {\rm p} \to {\rm D} + {\rm
e}^+ + \nu_{e})|^2}= G^2_{\rm V} g^2_{\rm A} M^4_{\rm
N}\,G^2_{\rm \pi NN}\,\frac{9Q_{\rm D}}{8\pi^2}\,F^2_{\rm D}(k^2)\nonumber\\
\hspace{-0.5in}&&\times\,\frac{\displaystyle C^2_0(k)}
{\displaystyle \Big[1 - \frac{1}{2}\,a^{\rm e}_{\rm pp}
r^{\rm e}_{\rm pp}k^2 + \frac{a^{\rm e}_{\rm pp}}{r_C}\,h(2 k r_C)\Big]^2 +
(a^{\rm e}_{\rm pp})^2 k^2 C^4_0(k)}\,\Bigg(-
g^{\alpha\beta}+\frac{k^{\alpha}_{\rm D}k^{\beta}_{\rm D}}{M^2_{\rm
D}}\Bigg)\nonumber\\ 
\hspace{-0.5in}&&\times {\rm tr}\{(- m_{\rm e} + \hat{k}_{\rm
e^+})\gamma_{\alpha}(1-\gamma^5) \hat{k}_{\nu_{\rm
e}}\gamma_{\beta}(1-\gamma^5)\}\times \frac{1}{4}\times {\rm tr}\{(M_{\rm
N} - \hat{p}_2) \gamma^5 (M_{\rm N} + \hat{p}_1) \gamma^5\},
\end{eqnarray}
where $m_{\rm e}=0.511\,{\rm MeV}$ is the positron mass, and we
have used the relation $g^2_{\rm V} = 2\,\pi^2\,Q_{\rm D}\,M^2_{\rm
N}$.

In the low--energy limit the computation of the traces yields
\begin{eqnarray}\label{label3.2}
\hspace{-0.5in}&&\Bigg(- g^{\alpha\beta}+\frac{k^{\alpha}_{\rm
D}k^{\beta}_{\rm D}}{M^2_{\rm D}}\Bigg)\,\times\,{\rm tr}\{( - m_{\rm e} +
\hat{k}_{\rm e^+})\gamma_{\alpha}(1-\gamma^5) \hat{k}_{\nu_{\rm
e}}\gamma_{\beta}(1-\gamma^5)\}= \nonumber\\
\hspace{-0.5in}&& = 24\,\Bigg( E_{\rm e^+} E_{\nu_{\rm e}} -
\frac{1}{3}\vec{k}_{\rm e^+}\cdot \vec{k}_{\nu_{\rm e}}\,\Bigg) ,\nonumber\\
\hspace{-0.5in}&&\frac{1}{4}\,\times\,{\rm tr}\{(M_{\rm N} - \hat{p}_2)
\gamma^5 (M_{\rm N} + \hat{p}_1) \gamma^5\} = 2\,M^2_{\rm N},
\end{eqnarray}
where we have neglected the relative kinetic energy of the protons with
respect to the mass of the proton.

Substituting Eq.~(\ref{label3.2}) in Eq.~(\ref{label3.1}) we get
\begin{eqnarray}\label{label3.3}
\hspace{-0.2in}&&\overline{|{\cal M}({\rm p} + {\rm p} \to {\rm D} + {\rm e}^+ +
\nu_{e})|^2} = \,G^2_{\rm V}\, g^2_{\rm A} M^6_{\rm N}\,G^2_{\rm
\pi NN}\,\frac{54 Q_{\rm D}}{\pi^2}\,F^2_{\rm D}(k^2)\nonumber\\
\hspace{-0.2in}&&\times\,\frac{\displaystyle C^2_0(k)}
{\displaystyle \Big[1 - \frac{1}{2}\,a^{\rm e}_{\rm pp}
r^{\rm e}_{\rm pp}k^2 + \frac{a^{\rm e}_{\rm pp}}{r_C}\,h(2 k r_C)\Big]^2 +
(a^{\rm e}_{\rm pp})^2 k^2 C^4_0(k)}\,\Bigg( E_{\rm e^+} E_{\nu_{\rm e}} -
\frac{1}{3}\vec{k}_{\rm e^+}\cdot \vec{k}_{\nu_{\rm e}}\Bigg).
\end{eqnarray}
The integration over the phase volume of the final ${\rm D}{\rm e}^+
\nu_{\rm e}$--state we perform in the non--relativistic limit
\begin{eqnarray}\label{label3.4}
\hspace{-0.5in}&&\int\frac{d^3k_{\rm D}}{(2\pi)^3 2E_{\rm
D}}\frac{d^3k_{\rm e^+}}{(2\pi)^3 2E_{\rm e^+}}\frac{d^3k_{\nu_{\rm
e}}}{(2\pi)^3 2 E_{\nu_{\rm e}}}\,(2\pi)^4\,\delta^{(4)}(k_{\rm D} +
k_{\ell} - p_1 - p_2)\,\Bigg( E_{\rm e^+} E_{\nu_{\rm e}} -
\frac{1}{3}\vec{k}_{\rm e^+}\cdot \vec{k}_{\nu_{\rm e}}\,\Bigg)\nonumber\\
\hspace{-0.5in}&&= \frac{1}{32\pi^3 M_{\rm N}}\,\int^{W + T_{\rm
pp}}_{m_{\rm e}}\sqrt{E^2_{\rm e^+}-m^2_{\rm e}}\,E_{\rm e^+}(W + T_{\rm
pp} - E_{\rm e^+})^2\,d E_{\rm e^+} = \frac{(W + T_{\rm pp})^5}{960\pi^3
M_{\rm N}}\,f(\xi),
\end{eqnarray}
where $W = \varepsilon_{\rm D} - (M_{\rm n} - M_{\rm p}) = (2.225
-1.293)\,{\rm MeV} = 0.932\,{\rm MeV}$ and $\xi = m_{\rm e}/(W +
T_{\rm pp})$. The function $f(\xi)$ is defined by the integral
\begin{eqnarray}\label{label3.5}
\hspace{-0.5in}f(\xi)&=&30\,\int^1_{\xi}\sqrt{x^2 -\xi^2}\,x\,(1-x)^2 dx=(1
- \frac{9}{2}\,\xi^2 - 4\,\xi^4)\,\sqrt{1-\xi^2}\nonumber\\
\hspace{-0.5in}&&+ \frac{15}{2}\,\xi^4\,{\ell
n}\Bigg(\frac{1+\sqrt{1-\xi^2}}{\xi}\Bigg)\Bigg|_{T_{\rm pp} = 0} =  0.222
\end{eqnarray}
and normalized to unity at $\xi = 0$.

Thus, the cross section for the solar proton burning is given by
\begin{eqnarray}\label{label3.6}
&&\sigma_{\rm pp}(T_{\rm pp}) = \frac{e^{\displaystyle
- \pi/r_C\sqrt{M_{\rm N}T_{\rm pp}}}}{v^2}\,
\alpha\,\frac{9g^2_{\rm A} G^2_{\rm V} Q_{\rm D}
M^3_{\rm N}}{320\,\pi^4}\,G^2_{\rm \pi NN}\,(W + T_{\rm
pp})^5\,f\Bigg(\frac{m_{\rm e}}{W + T_{\rm pp}}\Bigg)\nonumber\\
&&\times\,\frac{\displaystyle F^2_{\rm D}(M_{\rm N}T_{\rm pp})}{
\displaystyle \Big[1 - \frac{1}{2}\,a^{\rm e}_{\rm pp}
r^{\rm e}_{\rm pp}M_{\rm N}T_{\rm pp} + \frac{a^{\rm e}_{\rm
pp}}{r_C}\,h(2 r_C\sqrt{M_{\rm N}T_{\rm pp}})\Big]^2 + (a^{\rm e}_{\rm
pp})^2M_{\rm N}T_{\rm pp} C^4_0(\sqrt{M_{\rm N}T_{\rm pp}})}
=\nonumber\\ &&\hspace{1.2in} = \frac{S_{\rm pp}(T_{\rm pp})}{T_{\rm
pp}}\,e^{\displaystyle - \pi/r_C\sqrt{M_{\rm N}T_{\rm pp}}}.
\end{eqnarray}
The astrophysical factor $S_{\rm pp}(T_{\rm pp})$ reads
\begin{eqnarray}\label{label3.7}
\hspace{-0.5in}&&S_{\rm pp}(T_{\rm pp}) = 
\alpha\,\frac{9g^2_{\rm A}G^2_{\rm V}Q_{\rm
D}M^4_{\rm N}} {1280\pi^4}\,G^2_{\rm \pi NN}\,(W + T_{\rm
pp})^5\,f\Bigg(\frac{m_{\rm e}}{W + T_{\rm pp}}\Bigg)\nonumber\\
\hspace{-0.5in}&&\times\, \frac{\displaystyle F^2_{\rm D}(M_{\rm
N}T_{\rm pp})} {\displaystyle
\Big[1 - \frac{1}{2}\,a^{\rm e}_{\rm pp} r^{\rm e}_{\rm pp}M_{\rm
N}T_{\rm pp} + \frac{a^{\rm e}_{\rm pp}}{r_C}\,h(2 r_C\sqrt{M_{\rm
N}T_{\rm pp}})\Big]^2 + (a^{\rm e}_{\rm pp})^2M_{\rm N}T_{\rm pp}
C^4_0(\sqrt{M_{\rm N}T_{\rm pp}})}.
\end{eqnarray}
At zero kinetic energy of the relative movement of the protons $T_{\rm
pp} = 0$ the astrophysical factor $S_{\rm pp}(0)$ is given by
\begin{eqnarray}\label{label3.8}
\hspace{-0.5in}S_{\rm pp}(0) =\alpha\,\frac{9g^2_{\rm A}G^2_{\rm V}Q_{\rm
D}M^4_{\rm N}}{1280\pi^4}\,G^2_{\rm \pi NN}\,W^5\,f\Bigg(\frac{m_{\rm
e}}{W}\Bigg) =  4.08\,\times 10^{-25}\,{\rm MeV\,\rm b}.
\end{eqnarray}
The value $S_{\rm pp}(0) = 4.08 \times 10^{-25}\,{\rm MeV\,\rm b}$
agrees good with the recommended value $S_{\rm pp}(0) = 4.00 \times
10^{-25}\,{\rm MeV\,\rm b}$ [4]. Insignificant disagreement with the
result obtained in Ref.[11] where we have found $S_{\rm pp}(0) = 4.02
\times 10^{-25}\,{\rm MeV\,\rm b}$ is due to the new value of the
constant $g_{\rm A} = 1.260 \to 1.267$ [18] (see Ref.[13]).

Unlike the astrophysical factor obtained by
Kamionkowski and Bahcall [5] the astrophysical factor given by
Eq.(\ref{label3.8}) does not depend explicitly on the S--wave
scattering wave of pp scattering. This is due to the normalization of
the wave function of the relative movement of two protons. After the
summation of an infinite series and by using the relation
Eq.(\ref{label2.19}) we obtain the wave function of two protons in the
form
\begin{eqnarray}\label{label3.9}
\psi_{\rm pp}(k)= e^{\textstyle i\,\delta^{\,\rm e}_{\rm
pp}(k)}\,\frac{\sin\,\delta^{\,\rm e}_{\rm pp}(k)}{-a^{\rm e}_{\rm pp}kC_0(k)},
\end{eqnarray}
that corresponds the normalization of the wave function of the
relative movement of two protons used by Schiavilla {\it et al.}
[6]. For the more detailed discussion of this problem we relegate
readers to the paper by Schiavilla {\it et al.} [6]\footnote{See the
last paragraph of Sect.\,3 and the first paragraph of Sect.\,5 of
Ref.[6].}.

Unfortunately, the value of the astrophysical factor $S_{\rm pp}(0) =
4.08 \times 10^{-25}\,{\rm MeV\,\rm b}$ does not confirm the
enhancement by a factor of 1.4 obtained in the modified version of the
RFMD in Ref.[14].

\section{Neutrino disintegration of the deuteron induced
 by charged weak current}
\setcounter{equation}{0}

The evaluation of the amplitude of the process $\nu_{\rm e}$ + D $\to$
${\rm e}^-$ + p + p has been given in details in Ref.\,[10]. The result
can be written in the following form
\begin{eqnarray}\label{label4.1}
\hspace{-0.3in}&& i{\cal M}({\rm p} + {\rm p} \to {\rm D} + {\rm e}^+
+ \nu_{e}) = g_{\rm A} M_{\rm N} \frac{G_{\rm
V}}{\sqrt{2}}\,\frac{3g_{\rm V}}{2\pi^2}\, G_{\rm \pi NN}
\,e^*_{\mu}(k_{\rm D})\,[\bar{u}(k_{\rm e^-})\gamma^{\mu}(1-\gamma^5)
u(k_{\nu_{\rm e}})]\,{\cal F}^{\rm e}_{\rm
ppe^-}\nonumber\\\hspace{-0.3in} &&\times
\,\frac{C_0(k)}{\displaystyle1 - \frac{1}{2}\,a^{\rm e}_{\rm pp}
r^{\rm e}_{\rm pp}k^2 + \frac{a^{\rm e}_{\rm pp}}{r_C}\,h(2 k r_C) +
i\,a^{\rm e}_{\rm pp}\,k\,C^2_0(k) }\,[\bar{u^c}(p_2) \gamma^5
u(p_1)]\,F_{\rm D}(k^2),
\end{eqnarray}
where ${\cal F}^{\rm e}_{\rm ppe^-}$ is the overlap
factor which we evaluate below, and $F_{\rm D}(k^2)$ is the
universal form factor Eq.(\ref{label2.17}) describing a spatial
smearing of the deuteron [14].

The amplitude Eq.(\ref{label4.1}) squared, averaged over
polarizations of the deuteron and summed over polarizations of the
final particles reads
\begin{eqnarray}\label{label4.2}
&&\overline{|{\cal M}(\nu_{\rm e} + {\rm D} \to {\rm e}^- + {\rm p} +
{\rm p})|^2} = g^2_{\rm A}M^6_{\rm N}\frac{144 G^2_{\rm V}Q_{\rm
D}}{\pi^2}\,G^2_{\rm \pi NN}\,|{\cal F}^{\rm e}_{\rm
ppe^-}|^2\,\,F^2_{\rm D}(k^2)\,F(Z, E_{\rm e^-})\nonumber\\ &&\times
{\displaystyle \frac{\displaystyle C^2_0(k)}{\displaystyle \Big[1 -
\frac{1}{2}a^{\rm e}_{\rm pp} r^{\rm e}_{\rm pp} k^2 + \frac{a^{\rm
e}_{\rm pp}}{r_{\rm C}}\,h(2kr_{\rm C})\Big]^2 + (a^{\rm e}_{\rm
pp})^2k^2 C^4_0(k)}} \Bigg( E_{{\rm e^-}}E_{{\nu}_{\rm e}} -
\frac{1}{3}\vec{k}_{{\rm e^-}}\cdot \vec{k}_{{\nu}_{\rm e}}\Bigg),
\end{eqnarray}
where $F(Z,E_{\rm e^-}$ is the Fermi function [21] describing the
Coulomb interaction of the electron with the nuclear system having a
charge $Z$. In the case of the reaction $\nu_{\rm e}$ + D $\to$ e$^+$
+ p + p we have $Z = 2$. At $\alpha^2 Z^2 \ll 1$ the Fermi function
$F(Z,E_{\rm e^-})$ reads [21]
\begin{eqnarray}\label{label4.3}
F(Z,E_{\rm e^-}) = \frac{2\pi \eta_{\rm e^-}}{\displaystyle 1 -
e^{\textstyle -2\pi \eta_{\rm e^-}}},
\end{eqnarray}
where $\eta_{\rm e^-} = Z \alpha/v_{\rm e^-} = Z \alpha E_{\rm
e^-}/\sqrt{E^2_{\rm e^-} -m^2_{\rm e^-} }$ and $v_{\rm e^-}$ is a
velocity of the electron.

The r.h.s. of Eq.(\ref{label4.2}) can be expressed in terms of the
astrophysical factor $S_{\rm pp}(0)$ for the solar proton burning
brought up to the form
\begin{eqnarray}\label{label4.4}
\hspace{-0.5in}&&\overline{|{\cal M}(\nu_{\rm e} + {\rm D} \to {\rm
e}^- + {\rm p} + {\rm p})|^2} = S_{\rm
pp}(0)\,\frac{2^{12}5\pi^2}{\Omega_{\rm D e^+ \nu_{\rm
e}}}\,\frac{r_{\rm C}M^3_{\rm N}}{m^5_{\rm e}}\,\frac{|{\cal F}^{\rm e}_{\rm
ppe^-}|^2}{|{\cal F}^{\rm e}_{\rm
pp}|^2}\,F^2_{\rm
D}(k^2)\,F(Z, E_{\rm e^-})\nonumber\\
\hspace{-0.5in}&&\times \frac{\displaystyle C^2_0(k)}{\displaystyle
\Big[1 - \frac{1}{2}a^{\rm e}_{\rm pp} r^{\rm e}_{\rm pp} k^2 +
\frac{a^{\rm e}_{\rm pp}}{r_{\rm C}}\,h(2kr_{\rm C})\Big]^2 + (a^{\rm
e}_{\rm pp})^2k^2 C^4_0(k)}\, \Bigg( E_{{\rm e^-}}E_{{\nu}_{\rm e}} -
\frac{1}{3}\vec{k}_{{\rm e^-}} \cdot \vec{k}_{{\nu}_{\rm e}}\Bigg).
\end{eqnarray}
We have used here the expression for the astrophysical factor
\begin{eqnarray}\label{label4.5}
S_{\rm pp}(0) = \frac{9g^2_{\rm A}G^2_{\rm V}Q_{\rm D}M^3_{\rm
N}}{1280\pi^4r_{\rm C}}\,G^2_{\rm \pi NN}\,|{\cal F}^{\rm e}_{\rm
pp}|^2\,m^5_{\rm e}\,\Omega_{\rm D e^+ \nu_{\rm e}},
\end{eqnarray}
where $m_{\rm e} = 0.511\,{\rm MeV}$ is the electron mass, and
$\Omega_{\rm D e^+ \nu_{\rm e}} = (W/m_{\rm e})^5 f(m_{\rm e}/W) =
4.481$ at $W = 0.932\,{\rm MeV}$. The function $f(m_{\rm e}/W)$ is
defined by Eq.(\ref{label3.5}).

In the rest frame of the deuteron the cross section for the process
$\nu_{\rm e}$ + D $\to$ ${\rm e}^-$ + p + p is defined as
\begin{eqnarray}\label{label4.6}
&&\sigma^{\nu_{\rm e} D}_{\rm cc}(E_{\nu_{\rm e}}) = \frac{1}{4M_{\rm
D}E_{\nu_{\rm e}}}\int\,\overline{|{\cal M}(\nu_{\rm e} + {\rm D} \to
{\rm e}^- + {\rm p} + {\rm p})|^2}\nonumber\\
&&\frac{1}{2}\,(2\pi)^4\,\delta^{(4)}(k_{\rm D} + k_{\nu_{\rm e}} -
p_1 - p_2 - k_{\rm e^-})\, \frac{d^3p_1}{(2\pi)^3 2E_1}\frac{d^3
p_2}{(2\pi)^3 2E_2}\frac{d^3k_{{\rm e^-}}}{(2\pi)^3 2E_{{\rm e^-}}},
\end{eqnarray}
where $E_{\nu_{\rm e}}$, $E_1$, $E_2$ and $E_{{\rm e^-}}$ are the
energies of the neutrino, the protons and the electron. The
abbreviation (cc) means the charged current. The integration over the
phase volume of the (${\rm p p e^-}$)--state we perform in the
non--relativistic limit and in the rest frame of the deuteron,
\begin{eqnarray}\label{label4.7}
&&\frac{1}{2}\,\int\frac{d^3p_1}{(2\pi)^3 2E_1}\frac{d^3p_2}{(2\pi)^3
2E_2} \frac{d^3k_{\rm e}}{(2\pi)^3 2E_{\rm
e^-}}(2\pi)^4\,\delta^{(4)}(k_{\rm D} + k_{\nu_{\rm e}} - p_1 - p_2 -
k_{\rm e^-})\,\nonumber\\ &&{\displaystyle \frac{\displaystyle
C^2_0(\sqrt{M_{\rm N}T_{\rm pp}})\,F^2_{\rm D}(M_{\rm N}T_{\rm
pp})\,F(Z, E_{\rm e^-})}{\displaystyle \Big[1 - \frac{1}{2}a^{\rm
e}_{\rm pp} r^{\rm e}_{\rm pp}M_{\rm N}T_{\rm pp} + \frac{a^{\rm
e}_{\rm pp}}{r_{\rm C}}\, h(2 r_{\rm C}\sqrt{M_{\rm N}T_{\rm
pp}})\Big]^2 + (a^{\rm e}_{\rm pp})^2 M_{\rm N}T_{\rm pp}
C^4_0(\sqrt{M_{\rm N}T_{\rm pp}})}}\nonumber\\ &&\Bigg( E_{\rm e^-}
E_{\nu_{\rm e}} - \frac{1}{3} \vec{k}_{\rm e^-} \cdot
\vec{k}_{\nu_{\rm e}}\Bigg)\, =\frac{E_{\bar{\nu}_{\rm e}}M^3_{\rm
N}}{128\pi^3}\,\Bigg(\frac{E_{\rm th}}{M_{\rm N}}
\Bigg)^{\!\!7/2}\Bigg(\frac{2 m_{\rm e}}{E_{\rm
th}}\Bigg)^{\!\!3/2}\frac{1}{E^2_{\rm th}}\nonumber\\ &&\int\!\!\!\int
dT_{\rm e^-} dT_{\rm pp}\delta(E_{\nu_{\rm e}}- E_{\rm th} - T_{\rm
e^-} - T_{\rm pp}) \sqrt{T_{\rm e^-}T_{\rm pp}}\Bigg(1 + \frac{T_{\rm
e^-}}{m_{\rm e}}\Bigg)\,{\displaystyle \sqrt{1 + \frac{T_{\rm e^-}}{2
m_{\rm e}}}}\nonumber\\ &&{\displaystyle \frac{\displaystyle
C^2_0(\sqrt{M_{\rm N}T_{\rm pp}}) \,F^2_{\rm D}(M_{\rm N}T_{\rm
pp})\,F(Z, E_{\rm e^-})}{\displaystyle \Big[1 - \frac{1}{2}a^{\rm
e}_{\rm pp} r^{\rm e}_{\rm pp}M_{\rm N}T_{\rm pp} + \frac{a^{\rm
e}_{\rm pp}}{r_{\rm C}}\,h(2 r_{\rm C}\sqrt{M_{\rm N}T_{\rm
pp}})\Big]^2 + (a^{\rm e}_{\rm pp})^2 M_{\rm N}T_{\rm pp}
C^4_0(\sqrt{M_{\rm N}T_{\rm pp}})}} \nonumber\\ &&= \frac{E_{\nu_{\rm
e}}M^3_{\rm N}}{128\pi^3} \,\Bigg(\frac{E_{\rm th}}{M_{\rm N}}
\Bigg)^{\!\!7/2}\Bigg(\frac{2 m_{\rm e}}{E_{\rm
th}}\Bigg)^{\!\!3/2}\,(y-1)^2\,\Omega_{\rm p p e^-}(y),
\end{eqnarray}
where $T_{\rm e^-}$ is the kinetic energy of the electron, $E_{\rm th}$ is
the neutrino energy threshold of the reaction $\nu_{\rm e}$ + D $\to$ ${\rm
e}^-$ + p + p, and is given by $E_{\rm th}= \varepsilon_{\rm D} + m_{\rm e}
- (M_{\rm n} - M_{\rm p}) = (2.225 + 0.511 - 1.293) \, {\rm MeV} =
1.443\,{\rm MeV}$. The function $\Omega_{\rm p p e^-}(y)$, where
$y=E_{\nu_{\rm
e}}/E_{\rm th}$, is defined as
\begin{eqnarray}\label{label4.8}
\hspace{-0.5in}&&\Omega_{\rm p p e^-}(y) =  \int\limits^{1}_{0} dx
\sqrt{x (1 - x)} \Bigg(1 + \frac{E_{\rm th}}{m_{\rm
e}}(y-1)(1-x)\Bigg) \sqrt{1 + \frac{E_{\rm th}}{2 m_{\rm
e}}(y-1)(1-x)}\nonumber\\
\hspace{-0.5in}&&C^2_0(\sqrt{M_{\rm N}E_{\rm th}\,(y - 1)\,x})\,F^2_{\rm
D}(M_{\rm N}E_{\rm th}\,(y - 1)\,x)\,F(Z,m_{\rm e} + E_{\rm th}(y -
1)\,(1-x))\nonumber\\
\hspace{-0.5in}&&\Bigg\{\Bigg[1 - \frac{1}{2}a^{\rm e}_{\rm pp} r^{\rm
e}_{\rm pp}M_{\rm N}E_{\rm th}\,(y - 1)\,x + \frac{a^{\rm e}_{\rm
pp}}{r_{\rm C}}\,h(2 r_{\rm C}\sqrt{M_{\rm N}E_{\rm th}\,(y -
1)\,x})\Bigg]^2 \nonumber\\
\hspace{-0.5in}&&\hspace{0.2in} + (a^{\rm e}_{\rm pp})^2\,M_{\rm N}E_{\rm
th}\,(y - 1)\,x C^4_0(\sqrt{M_{\rm N}E_{\rm th}\,(y - 1)\,x})
\Bigg\}^{-1}\!\!\!\!\!,
\end{eqnarray}
where we have changed the variable $T_{\rm pp} = (E_{\nu_{\rm e}} -
E_{\rm th})\,x$.

The cross section for $\nu_{\rm e}$ + D $\to$ ${\rm e}^-$ + p + p is defined
\begin{eqnarray}\label{label4.9}
\hspace{-0.5in}\sigma^{\nu_{\rm e} D}_{\rm cc}(E_{\nu_{\rm e}}) &=&
S_{\rm pp}(0)\, \frac{640 r_{\rm C}}{\pi \Omega_{\rm D e^+\nu_{\rm
e}}}\Bigg(\frac{M_{\rm N}}{E_{\rm th}}\Bigg)^{3/2}\Bigg(\frac{E_{\rm
th}}{2m_{\rm e}}\Bigg)^{7/2}\frac{|{\cal F}^{\rm e}_{\rm
ppe^-}|^2}{|{\cal F}^{\rm e}_{\rm
pp}|^2}\,(y-1)^2\,\Omega_{\rm p p
e^-}(y)=\nonumber\\ 
&=&3.69\times 10^5\,S_{\rm
pp}(0)\,\frac{|{\cal F}^{\rm e}_{\rm
ppe^-}|^2}{|{\cal F}^{\rm e}_{\rm
pp}|^2}\,(y-1)^2\,\Omega_{\rm p p e^-}(y),
\end{eqnarray}
where $S_{\rm pp}(0)$ is measured in ${\rm MeV}\,{\rm cm}^2$. For
$S_{\rm pp}(0) = 4.08\times 10^{-49}\,{\rm MeV}\,{\rm cm}^2$
Eq.(\ref{label3.8}) the cross section $\sigma^{\nu_{\rm e} D}_{\rm
cc}(E_{\nu_{\rm e}})$ reads
\begin{eqnarray}\label{label4.10}
\sigma^{\nu_{\rm e} D}_{\rm cc}(E_{\nu_{\rm e}}) =
1.50\,\frac{|{\cal F}^{\rm e}_{\rm
ppe^-}|^2}{|{\cal F}^{\rm e}_{\rm
pp}|^2}\,(y-1)^2\,\Omega_{\rm p p e^-}(y)\,10^{-43}\,{\rm cm}^2.
\end{eqnarray}
In order to make numerical predictions for the cross section
Eq.(\ref{label4.10}) we should evaluate the overlap factor ${\cal
F}^{\rm e}_{\rm ppe^-}$.  This evaluation can be carried out in
analogy with the evaluation of ${\cal F}^{\rm e}_{\rm pp}$.  By using
the results obtained in Ref.[10] we get
\begin{eqnarray}\label{label4.11}
\hspace{-0.2in}{\cal F}^{\rm e}_{\rm ppe^-} = 1 +
\frac{32}{9}\int\limits^{\infty}_0 dv\,v^2\,\Bigg[e^{\textstyle -
\alpha\pi/4v}\,\Gamma\Bigg(1 - \frac{i\alpha}{2v}\Bigg)-
1\Bigg]\Bigg[\frac{1}{(1 + v^2)^{5/2}} - \frac{1}{16}\,\frac{1}{(1 +
v^2)^{3/2}}\Bigg].
\end{eqnarray}
The perturbative evaluation of the integral gives 
\begin{eqnarray}\label{label4.12}
{\cal F}^{\rm e}_{\rm ppe^-} = 1 -
\alpha\,\Bigg(\frac{13\pi}{54} - i\,\frac{13\gamma}{27}\Bigg) + O(\alpha^2).
\end{eqnarray}
Thus, the overlap factor ${\cal F}^{\rm e}_{\rm ppe^-}$ differs
slightly from unity as well as the overlap factor ${\cal F}^{\rm
e}_{\rm pp}$ of the solar proton burning. The ratio of the overlap
factors is equal to
\begin{eqnarray}\label{label4.13}
\frac{|{\cal F}^{\rm e}_{\rm ppe^-}|^2}{|{\cal F}^{\rm e}_{\rm pp}|^2}
= 1 - \alpha\,\frac{2\pi}{3} + O(\alpha^2) = 1 +
(-1.53 \times 10^{-2}) \simeq 1.
\end{eqnarray}
Setting $|{\cal F}^{\rm e}_{\rm ppe^-}|^2/|{\cal F}^{\rm e}_{\rm
pp}|^2 = 1$ we can make numerical predictions for the cross section
Eq.(\ref{label4.10}) and compare them with the PMA ones.  

The most recent PMA calculations the cross section for the reaction
$\nu_{\rm e}$ + D $\to$ e$^-$ + p + p have been obtained in
Refs.\,[22,23] and tabulated for the neutrino energies ranging over
the region from threshold up to 160$\,{\rm MeV}$. Since our result is
restricted by the neutrino energies from threshold up to 10$\,{\rm
MeV}$, we compute the cross section only for this energy region
\begin{eqnarray}\label{label4.14}
\sigma^{\nu_{\rm e} D}_{\rm cc}(E_{\nu_{\rm e}}= 4\,{\rm MeV}) &=&
2.46\,(1.86/1.54)\times 10^{-43}\,{\rm cm}^2,\nonumber\\
\sigma^{\nu_{\rm e} D}_{\rm cc}(E_{\nu_{\rm e}}= 6\,{\rm MeV}) &=&
9.60\,(5.89/6.13)\times 10^{-43}\,{\rm cm}^2,\nonumber\\
\sigma^{\nu_{\rm e} D}_{\rm cc}(E_{\nu_{\rm e}}= 8\,{\rm MeV}) &=&
2.38\,(1.38/1.44)\times 10^{-42}\,{\rm cm}^2,\nonumber\\
\sigma^{\nu_{\rm e} D}_{\rm cc}(E_{\nu_{\rm e}}= 10\,{\rm MeV}) &=&
4.07\,(2.55/2.66)\times 10^{-43}\,{\rm cm}^2,
\end{eqnarray}
where the data in parentheses are taken from Refs.\,[22] and [23],
respectively. Thus, on the average our numerical values for the cross
section $\sigma^{\nu_{\rm e} D}_{\rm cc}(E_{\nu_{\rm e}})$ by a factor
of 1.5 are larger compared with the PMA ones.

Our predictions for the cross section Eq.(\ref{label4.14}) differ from
the predictions of Ref.[14]. This is related to (i) the value of the
astrophysical factor which is by a factor 1.4 larger in Ref.[14] and
(ii) the form factor describing a spatial smearing of the deuteron
which is $F^2_{\rm D}(k^2)$ is this paper (see Ref. [13]) and $F_{\rm
D}(k^2)$ in Ref.[14]. 

\section{Astrophysical factor for pep process}
\setcounter{equation}{0}

In the RFMD the amplitude of the reaction p + e$^-$ + p $\to$ D +
$\nu_{\rm e}$ or the pep--process is related to the effective
Lagrangian Eq.(\ref{label2.3}) and reads
\begin{eqnarray}\label{label5.1}
&& i{\cal M}({\rm p} + {\rm e}^- + {\rm p} \to {\rm D} + \nu_{e}) =
G_{\rm V}\,g_{\rm A} M_{\rm N}\,G_{\rm \pi NN}\,\frac{3g_{\rm
V}}{4\pi^2}\,e^*_{\mu}(k_{\rm D})\,[\bar{u}(k_{\nu_{\rm
e}})\gamma^{\mu} (1-\gamma^5) u(k_{\rm e^-})]\,{\cal F}^{\rm e}_{\rm
pp}\nonumber\\ &&\times \,\frac{\displaystyle C_0(k)}{\displaystyle1 -
\frac{1}{2}\,a^{\rm e}_{\rm pp} r^{\rm e}_{\rm pp}k^2 + \frac{a^{\rm
e}_{\rm pp}}{r_C}\,h(2 k r_C) + i\,a^{\rm e}_{\rm pp}\,k\,C^2_0(k)
}\,[\bar{u^c}(p_2) \gamma^5 u(p_1)]\,F_{\rm D}(k^2),
\end{eqnarray}
where we have described low--energy elastic pp scattering in analogy
with the solar proton burning and the neutrino disintegration of the
deuteron.  

The amplitude Eq.(\ref{label5.1}) squared, averaged and
summed over polarizations of the interacting particles is defined
\begin{eqnarray}\label{label5.2}
&&\overline{|{\cal M}({\rm p} + {\rm e}^- + {\rm p} \to {\rm D} +
\nu_{e})|^2} = G^2_{\rm V}\, g^2_{\rm A} M^6_{\rm N}\,G^2_{\rm \pi
NN}\,\frac{27 Q_{\rm D}}{\pi^2}\,|{\cal F}^{\rm e}_{\rm
pp}|^2\,\,F^2_{\rm D}(k^2)\,F(Z, E_{\rm e^-})\nonumber\\
&&\times\,\frac{\displaystyle C^2_0(k)} {\displaystyle \Big[1 -
\frac{1}{2}\,a^{\rm e}_{\rm pp} r^{\rm e}_{\rm pp}k^2 + \frac{a^{\rm
e}_{\rm pp}}{r_C}\,h(2 k r_C)\Big]^2 + (a^{\rm e}_{\rm pp})^2 k^2
C^4_0(k)}\,\Bigg( E_{\rm e^+} E_{\nu_{\rm e}} -
\frac{1}{3}\vec{k}_{\rm e^+}\cdot \vec{k}_{\nu_{\rm e}}\Bigg),
\end{eqnarray}
where $F(Z, E_{\rm e^-})$ is the Fermi function given by
Eq.(\ref{label4.3}).

  At low energies the cross section $\sigma_{\rm pep}(T_{\rm pp})$ for
the pep--process can be determined as follows [24]
\begin{eqnarray}\label{label5.3}
&&\sigma_{\rm pep}(T_{\rm pp}) = \frac{1}{v}\frac{1}{4M^2_{\rm N}}\int
\frac{d^3k_{\rm e^-}}{(2\pi)^3 2 E_{\rm e^-}}\,g\, n(\vec{k}_{\rm
e^-})\int \overline{|{\cal M}({\rm p} + {\rm e}^- + {\rm p} \to {\rm
D} + \nu_{\rm e})|^2}\nonumber\\ &&(2\pi)^4 \delta^{(4)}(k_{\rm D} +
k_{\nu_{\rm e}} - p_1 - p_2 - k_{\rm e^-}) \frac{d^3k_{\rm
D}}{(2\pi)^3 2M_{\rm D}}\frac{d^3k_{\nu_{\rm e}}}{(2\pi)^3
2E_{\nu_{\rm e}}},
\end{eqnarray}
where $g = 2$ is the number of the electron spin states and $v$ is a
relative velocity of the protons. The electron distribution function
$n(\vec{k}_{\rm e^-})$ can be taken in the form [21]
\begin{eqnarray}\label{label5.4}
n(\vec{k}_{\rm e^-}) = e^{\displaystyle \bar{\nu} - T_{\rm e^-}/kT_c},
\end{eqnarray}
where $k = 8.617\times 10^{-11}\,{\rm MeV\,K^{-1}}$, $T_c$ is a
temperature of the core of the Sun.  The distribution function
$n(\vec{k}_{\rm e^-})$ is normalized by the condition
\begin{eqnarray}\label{label5.5}
g\int \frac{d^3k_{\rm e^-}}{(2\pi)^3}\,n(\vec{k}_{\rm e^-}) = n_{\rm e^-},
\end{eqnarray}
where $n_{\rm e^-}$ is the electron number density. From the
normalization condition Eq.(\ref{label5.5}) we derive
\begin{eqnarray}\label{label5.6}
e^{\displaystyle \bar{\nu}} = \frac{\displaystyle  4\,\pi^3\, n_{\rm
e^-}}{\displaystyle (2\pi\,m_{\rm e}\,kT_c)^{3/2}}.
\end{eqnarray}
The astrophysical factor $S_{\rm pep}(0)$ is then defined
by
\begin{eqnarray}\label{label5.7}
S_{\rm pep}(0) = S_{\rm pp}(0)\,\frac{15}{2\pi}\,\frac{1}{\Omega_{\rm D
e^+ \nu_{\rm e}}}\,\frac{1}{m^3_{\rm e}}\,\Bigg(\frac{E_{\rm
th}}{m_{\rm e}}\Bigg)^2\,e^{\displaystyle \bar{\nu}}\,\int d^3k_{\rm
e^-} \,e^{\displaystyle - T_{\rm e^-}/kT_c}\,F(Z, E_{\rm e^-}).
\end{eqnarray}
For the ratio $S_{\rm pep}(0)/S_{\rm pp}(0)$ we obtain
\begin{eqnarray}\label{label5.8}
\frac{S_{\rm pep}(0)}{S_{\rm pp}(0)} = \frac{2^{3/2}\pi^{5/2}}{f_{\rm
pp}(0)}\,\Bigg(\frac{\alpha Z n_{\rm e^-}}{m^3_{\rm
e}}\Bigg)\,\Bigg(\frac{E_{\rm th}}{m_{\rm
e}}\Bigg)^2\,\sqrt{\frac{m_{\rm e}}{k T_c}}\,I\Bigg(Z\sqrt{\frac{2
m_{\rm e}}{k T_c}}\Bigg).
\end{eqnarray}
We have set $f_{\rm pp}(0) = \Omega_{\rm D e^+ \nu_{\rm e}}/30 =
0.149$ [21] and the function $I(x)$ having been introduced by Bahcall
and May [21] reads
\begin{eqnarray}\label{label5.9}
I(x) = \int\limits^{\infty}_0 {\displaystyle \frac{\displaystyle
du\,e^{\displaystyle -u}}{\displaystyle 1 - e^{\displaystyle
-\pi\alpha\,x/\sqrt{u}}}}.
\end{eqnarray}
The relation between the astrophysical factors $S_{\rm pep}(0)$ and
$S_{\rm pp}(0)$ given by Eq.(\ref{label5.8}) is in complete agreement
with that obtained by Bahcall and May [21]. The ratio
Eq.(\ref{label5.8}) does not depend on whether the astrophysical
factor $S_{\rm pp}(0)$ is enhanced with respect to the recommended
value or not.

\section{Conclusion}
\setcounter{equation}{0}

We have shown that the contributions of low--energy elastic pp
scattering in the ${^1}{\rm S}_0$--state with the Coulomb repulsion to
the amplitudes of the reactions p + p $\to$ D + e$^+$ + $\nu_{\rm e}$,
$\nu_{\rm e}$ + D $\to$ e$^-$ + p + p and p + e$^-$ + p $\to$ D +
$\nu_{\rm e}$ can be described in the RFMD in full agreement with
low--energy nuclear phenomenology in terms of the S--wave scattering
length and the effective range. The amplitude of low--energy elastic
pp scattering has been obtained by summing up an infinite series of
one--proton loop diagrams and the evaluation of the result of the
summation in leading order in the large $N_C$ expansion. This takes
away fully the problem pointed out by Bahcall and Kamionkowski [17]
that in the RFMD with the effective local four--nucleon interaction
Eq.(\ref{label2.1}) one cannot describe low--energy elastic pp
scattering in the ${^1}{\rm S}_0$--state with the Coulomb repulsion in
agreement with low--energy nuclear phenomenology.

The obtained numerical value of the astrophysical factor $S_{\rm
pp}(0) = 4.08\times 10^{-25}\,{\rm MeV\, b}$ agrees with the
recommended value $S_{\rm pp}(0) = 4.00\times 10^{-25}\,{\rm MeV\, b}$
and recent estimate $S_{\rm pp}(0) = 4.20\times 10^{-25}\,{\rm MeV\,
b}$ [9] obtained from the helioseismic data. 

Unfortunately, the value of the astrophysical factor $S_{\rm pp}(0) =
4.08 \times 10^{-25}\,{\rm MeV\,\rm b}$ does not confirm the
enhancement by a factor of 1.4 obtained in the modified version of the
RFMD in Ref.[14] which is not well defined due to a violation of
Lorentz invariance of the effective four--nucleon interaction
describing N + N $\to$ N + N transitions. This violation has turned
out to be incompatible with a dominance of one--nucleon loop anomalies
which are Lorentz covariant.

The cross section for the neutrino disintegration of the deuteron has
been evaluated with respect to $S_{\rm pp}(0)$. We have obtained an
enhancement of the cross section by a factor of order 1.5 on the
average for neutrino energies $E_{\nu_{\rm e}}$ varying from threshold
to $E_{\nu_{\rm e}} \le 10\,{\rm MeV}$.  It would be important to
verify our results for the reaction $\nu_{\rm e}$ + D $\to$ e$^-$ + p
+ p in solar neutrino experiments planned by SNO. In fact, first, this
should provide an experimental study of $S_{\rm pp}(0)$ and, second,
the cross sections for the anti--neutrino disintegration of the
deuteron caused by charged $\bar{\nu}_{\rm e}$ + D $\to$ e$^+$ + n + n
and neutral $\bar{\nu}_{\rm e}$ + D $\to$ $\bar{\nu}_{\rm e}$ + n + p
weak currents have been found in good agreement with recent
experimental data obtained by the Reines's experimental group [26].

The evaluation of the astrophysical factor $S_{\rm pep}(0)$ for the
reaction p + e$^-$ + p $\to$ D + $\nu_{\rm e}$ or pep--process in the
RFMD has shown that the ratio $S_{\rm pep}(0)/S_{\rm pp}(0)$, first,
agrees fully with the result obtained by Bahcall and May [21] and,
second, does not depend on whether $S_{\rm pp}(0)$ is enhanced with
respect to the recommended value or not.

Concluding the paper we would like to emphasize that our model, the
RFMD, conveys the idea of a dominant role of one--fermion loop
(one--nucleon loop) anomalies from elementary particle physics to the nuclear
one. This is a new approach to the description of low--energy
nuclear forces in physics of finite nuclei. In spite of almost 30
year's history after the discovery of one--fermion loop anomalies and
application of these anomalies to the evaluation of effective
Lagrangians of low--energy interactions of hadrons, in nuclear physics
fermion--loop anomalies have not been applied to the analysis of
low--energy nuclear interactions and  properties of
nuclei. However, an important role of $N\bar{N}$ fluctuations for the
correct description of low--energy properties of finite nuclei has
been understood in Ref.[16]. Moreover, $N\bar{N}$ fluctuations have
been described in terms of one--nucleon loop diagrams within quantum
field theoretic approaches, but the contributions of one--nucleon loop
anomalies have not been considered in the papers of Ref.[16].

The RFMD strives to fill this blank. Within the framework of the RFMD
we aim to understand, in principle, the possibility of the description
of strong low-energy nuclear forces in terms of one--nucleon loop
anomalies. Of course, our results should be quantitatively compared
with the experimental data and other theoretical
approaches. Nevertheless, at the present level of the development of
our model one cannot demand at once to describe, for example, the
astrophysical factor $S_{\rm pp}(0)$ with accuracy better than it has
been carried out by Schiavilla {\it et al.} [6], where only
corrections not greater than 1$\%$ are allowed. It is not important
for our approach at present. What is much more important is in the
possibility to describe without free parameters in quantitative
agreement with both the experimental data and other theoretical
approaches all multitude of low--energy nuclear reactions of the
deuteron coupled to nucleons and other particles.  In Ref.[13] we have
outlined the procedure of the evaluation of chiral meson--loop
corrections in the RFMD.  The absence of free parameters in the RFMD
gives the possibility to value not only the role of these corrections
but also the corrections of other kind mentioned recently by Vogel and
Beacom [25].

The justification of the RFMD within QCD and large $N_C$ expansion
[12] implies that one--nucleon loop anomalies might be natural objects
for the understanding of low-energy nuclear forces. The real accuracy
of the approach should be found out for the process of the
development.

\section{Acknowledgement}

We would like to thank Prof. Kamionkowski and Prof. Beacom for reading
manuscript and useful comments.

\end{document}